\documentclass[aps,prd,groupedaddress,eqsecnum]{revtex4}
\usepackage{graphicx}
\newcommand\eqn[1]{(\ref{#1})}      

\newcommand{\be}{\begin{equation}}
\newcommand{\ee}{\end{equation}\noindent}
\newcommand{\bea}{\begin{eqnarray}}
\newcommand{\eea}{\end{eqnarray}\noindent}

\begin{document}
\title{New Strong-Field QED Effects at ELI: Nonperturbative Vacuum Pair Production}
\author{Gerald V.\ Dunne
%
}                     
%
%
\affiliation{Department of Physics, University of Connecticut, Storrs CT 06269-3046, USA}
%
%
\begin{abstract}
Since the work of Sauter, and Heisenberg, Euler and K\"ockel, it has been understood that vacuum polarization effects in quantum electrodynamics (QED) predict remarkable new phenomena such as light-light scattering and pair production from vacuum. However, these fundamental effects are difficult to probe experimentally because they are very weak, and they are difficult to analyze theoretically because they are highly nonlinear and/or nonperturbative. The Extreme Light Infrastructure (ELI) project offers the possibility of a new window into this largely unexplored world. I review these ideas, along with some new results, explaining why quantum field theorists are so interested in this rapidly developing field of laser science. I concentrate on the theoretical tools that have been developed to analyze  nonperturbative vacuum pair production.
%
%
\end{abstract}
\maketitle
\section{Introduction}
\label{intro}
The Extreme Light Infrastructure (ELI) project \cite{eli} will provide lasers with electromagnetic fields of unprecedented intensity, allowing a new experimental window into the largely unexplored  regime of nonperturbative quantum electrodynamics (QED). This has implications not just for QED, but also for fundamental issues in quantum field theory, as well as nuclear, atomic, plasma, gravitational and astro- physics. 
Quantum vacuum fluctuations mean that the QED vacuum behaves like a polarizable medium that modifies classical behavior, leading to novel quantum effects \cite{he,gvd,viki1,greiner,dittrichgies}. Many of these, such as the Lamb shift \cite{lamb,karshenboim}, Delbr\"uck scattering \cite{akh-del}, photon splitting \cite{adler,akh-splitting}, nonlinear Compton scattering \cite{compton}, and the Casimir effect \cite{casimir,jaffe,bordag}, have been experimentally observed, while others such as elastic photon-photon scattering and nonperturbative electron-positron pair production from vacuum, have not yet been observed.  This is essentially because they are very weak effects. Our current understanding of the {\it perturbative} regime of QED is extremely good. The classic example is the anomalous magnetic moment $g$ of the electron [defined by the proportionality between magnetic moment and spin: $\vec{\mu}=g \frac{e}{2m} \vec{S}$], concerning which there has been dramatic recent progress, both experimentally and theoretically. On the theoretical side \cite{kinoshita}, a four-loop computation [involving 891 four-loop Feynman diagrams!]  gives $g$ as a series in the fine structure constant $\alpha=\frac{e^2}{\hbar c}$ [which sets the interaction strength of perturbative QED] as:
\begin{eqnarray}
\left(\frac{g-2}{2}\right)\Bigg|_{\rm th}&=&\frac{1}{2}\, \frac{\alpha}{\pi}-0.32848... \left(\frac{\alpha}{\pi}\right)^2 +
1.18124...  \left(\frac{\alpha}{\pi}\right)^3 -
1.9144(35) \left(\frac{\alpha}{\pi}\right)^4+...
\label{g-2}
\end{eqnarray}
On the experimental side \cite{gabrielse},  a single-electron cyclotron has been used to measure $g$ directly, without measuring $\alpha$, and the result agrees remarkably well  with the value obtained from \eqn{g-2} using a precise  independent measurement of $\alpha$ (from atomic recoil experiments with Rb):
\begin{eqnarray}
\left(\frac{g-2}{2}\right)\Bigg|_{\rm exp}&=&0.001\,159\,652\,180\,73\,(28)\nonumber\\
\left(\frac{g-2}{2}\right)\Bigg|_{\rm Rb}&=&0.001\,159\,652\,178\,86
\label{comp}
\end{eqnarray}
This is an impressive confirmation of the precision of perturbative QED.

By contrast, we know very little about the nonperturbative regime of QED that arises when we consider QED in ultra-strong external fields. To quantify  "ultra-strong", recall the computation of Heisenberg and Euler \cite{he} of the probability of vacuum pair production in an applied uniform electric field of strength ${\mathcal E}$. Vacuum pair production is a process in which virtual dipole pairs in the vacuum can be accelerated apart by the external field, becoming real asymptotic $e^+\,e^-$ pairs if they gain the binding energy of $2mc^2$ from the external field, as depicted in Figure \ref{fig1}. This is a non-perturbative process, and the leading exponential part of the probability, assuming a constant electric field, was computed  by Heisenberg and Euler \cite{he,gvd}:
\begin{figure}
\centerline{\resizebox{0.4\textwidth}{!}{%
\includegraphics{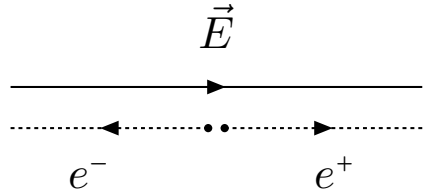}}
}
\caption{Pair production as the separation of a virtual vacuum dipole pair under the influence of an external electric field.}
\label{fig1}
\end{figure}
\begin{equation}
P_{\rm HE} \sim  \exp\left[{-\frac{\pi\, m^2\,c^3}{e\, {\mathcal E} \,\hbar}}\right]\quad ,
 \label{constant}
 \end{equation}
 building on earlier work of Sauter \cite{sauter}.
This result sets a basic scale of a critical field strength and intensity near which we expect to observe such nonperturbative effects:
\begin{eqnarray}
 {\mathcal E}_{\rm 
c}&=&\frac{m^2 c^3}{e\,\hbar}\approx  10^{16}\,\,{\rm V/cm} \nonumber\\ I_c&=&\frac{c}{8\pi}\,{\mathcal E}^2_c\approx 4\times 10^{29}\,\, {\rm W/cm^2}
\label{critical}
\end{eqnarray}

As a useful guiding analogy, recall Oppenheimer's computation \cite{opp} of the probability of ionization of an atom of binding energy $E_b$ in such a uniform electric field:
\begin{equation}
P_{\rm ionization}\sim \exp\left[-\frac{4}{3}\frac{\sqrt{2m}\,
E_b^{3/2}}{e{ \mathcal
E}\hbar}\right] \quad .
 \label{oppie}
 \end{equation}
Taking as a representative atomic energy scale the binding energy of hydrogen, $E_b=\frac{m e^4}{2\hbar^2}\approx 13.6 \, {\rm eV}$,  we find
\begin{equation}
P^{\rm hydrogen}\sim \exp\left[ -\frac{2}{3} \frac{m^2\, e^5}{{\mathcal E}\, \hbar^4}\right]\quad .
\label{hydrogen}
\end{equation}
This result sets a basic scale of field strength and intensity near which we expect to observe such nonperturbative ionization effects in atomic systems:
\begin{eqnarray}
 {\mathcal E}_{\rm 
c}^{\rm ionization}&=&\frac{m^2 e^5}{\hbar^4}=\alpha^3  {\mathcal E}_{\rm 
c} \approx  4\times 10^{9}\,\,{\rm V/cm} \nonumber\\
I_c^{\rm ionization}&=&\alpha^6 I_c \approx 6\times 10^{16}\,\, {\rm W/cm^2}
\end{eqnarray}
These, indeed,  are the familiar scales of atomic ionization experiments. Note that ${\mathcal E}_{\rm 
c}^{\rm ionization}$ differs from ${\mathcal E}_{\rm 
c}$ by  a factor of $\alpha^3\sim 4\times 10^{-7}$.
These simple estimates explain why vacuum pair production has not yet been observed -- it is an astonishingly  weak effect with conventional lasers \cite{bunkin,ringwald}. This is because it is primarily a non-perturbative effect, that depends exponentially on the (inverse) electric field strength, and there is a factor of $\sim 10^7$ difference between the critical field scales in the atomic regime and in the vacuum pair production regime. Thus,  with standard lasers that can routinely probe ionization, there is no hope to see vacuum pair production. However, recent technological advances in laser science, and also in theoretical refinements of the Heisenberg-Euler computation, suggest that lasers such as those planned for ELI may be able to reach this elusive nonperturbative regime. This has the potential to open up an entirely new domain of experiments, with the prospect of fundamental discoveries and practical applications, as are described in many talks in this conference.

\section{The QED Effective Action}

In quantum field theory, the key object that encodes vacuum polarization corrections to classical Maxwell electrodynamics is the "effective action" $\Gamma[A]$, which is a functional of the applied classical gauge field $A_\mu(x)$ \cite{schwinger-qed,schwinger,dittrichreuter}. The effective action is the relativistic quantum field theory analogue of the grand potential of statistical physics, in the sense that it contains a wealth of information about the quantum system: here, the nonlinear properties of the quantum vacuum. For example, the polarization tensor $\Pi_{\mu\nu}=\frac{\delta^2\Gamma}{\delta A_\mu \delta A_\nu}$ contains the electric permittivity $\epsilon_{ij}$ and the magnetic permeability $\mu_{ij}$ of the quantum vacuum, and is obtained by varying the effective action $\Gamma[A]$ with respect to the external probe $A_\mu(x)$. The general formalism for the QED effective action was developed in a series of papers by Schwinger in the 1950's \cite{schwinger-qed,schwinger}. $\Gamma[A]$ is defined \cite{schwinger} in terms of the vacuum-vacuum persistence amplitude 
\begin{equation}
\langle 0_{\rm out} \, | \, 0_{\rm in} \rangle =\exp\left[\frac{i}{\hbar}\left\{ {\rm Re}(\Gamma)+i\, {\rm Im}(\Gamma)\right\}\right]
\label{persistence}
\end{equation}
Note that $\Gamma[A]$ has a real part that describes dispersive effects such as vacuum birefringence, and an imaginary part that describes absorptive effects, such as vacuum pair production. Dispersive effects are discussed in detail in Holger Gies's contribution to this Volume.
The imaginary part encodes the probability of vacuum pair production as 
\begin{eqnarray}
P_{\rm production}&=&1-|\langle 0_{\rm out} \, | \, 0_{\rm in} \rangle |^2\nonumber\\
&=& 1-\exp\left[-\frac{2}{\hbar}\, {\rm Im}\, \Gamma\right]\nonumber\\
&\approx& \frac{2}{\hbar}\, {\rm Im}\, \Gamma
\label{prob}
\end{eqnarray}
Here, in the last [approximate] step we use the fact that ${\rm Im}(\Gamma)/\hbar$ is typically very small. The expression (\ref{prob}) can be viewed as the relativistic quantum field theoretic analogue of the well-known quantum mechanical fact that the ionization probability is determined by the imaginary part of the energy of an atomic electron in an applied electric field.

From a computational perspective, the effective action is defined as \cite{schwinger-qed,schwinger,dittrichreuter}
\begin{eqnarray}
\Gamma[A]&=&\hbar \, \ln\, \det\left[ i D\hskip -7pt /  -m\right]\nonumber\\
&=& \hbar \, {\rm tr} \, \ln \left[ i D\hskip -7pt /  -m\right]  \quad .
\end{eqnarray}
Here, $D\hskip -7pt / \equiv \gamma^\mu D_\mu$, with $\gamma_\mu$ being the Dirac gamma matrices of relativistic quantum mechanics, and the covariant derivative operator,  $D_\mu=\partial_\mu-i\frac{e}{\hbar c}A_\mu$, defines the coupling between electrons and the electromagnetic field $A_\mu$.
When the gauge field $A_\mu$ is such that the field strength, $F_{\mu\nu}=\partial_\mu A_\nu-\partial_\nu A_\mu$, is constant, the spectral problem becomes that of 2 harmonic oscillators, as is easily seen in the Fock-Schwinger gauge: $A_\mu=-\frac{1}{2}F_{\mu\nu} x^\nu$. In this case, $F_{\mu\nu}$ is a $4\times 4$ antisymmetric matrix, so its eigenvalues come in $\pm$ pairs, and can be expressed simply in terms of the two relativistic invariants $(\vec{E}^2-\vec{B}^2)$ and $\vec{E}\cdot\vec{B}$. Then the determinant can be computed in closed form \cite{he,gvd}. This is like a relativistic analogue of Landau's famous computation of diamagnetism for nonrelativistic electrons in a  constant magnetic field \cite{landau-1}. For a constant {\it electric} field, of strength  ${\mathcal E}$, the result of Heisenberg and Euler is
\begin{equation}
\Gamma^{\rm HE}=-\hbar\,{\rm Vol_4}\frac{e^2 {\mathcal E}^2}{8\pi^2}\int_0^\infty \frac{ds}{s^2}\, e^{-\frac{m^2}{e{\mathcal E}}s}\left(\cot(s)-\frac{1}{s}+\frac{s}{3}\right)
\label{hel}
\end{equation}
The poles of ${\rm cot(s)}$ lead to an imaginary part of $\Gamma$, whose leading behavior is determined by the first pole at $s=\pi$:
\begin{equation}
 \frac{{\rm  Im}\, \Gamma^{\rm HE}}{\rm Vol_4} \sim \hbar \frac{e^2\,{\mathcal E}^2}{8\pi^3} \exp\left[-\frac{\pi\, m^2}{e\, {\mathcal E}}\right]
 \label{constant-2}
 \end{equation}
(we now work in units with  $\hbar=c=1$).
This result, and its interpretation in terms of vacuum instability, was stated clearly already by Heisenberg and Euler \cite{he}, building on earlier ideas of Sauter \cite{sauter}.
 
 There is an interesting interpretation of this result as a virial expansion \cite{nikishov-review,ritus-review,lebedev}, in which we write
 \begin{equation}
 \frac{2}{\hbar}\, {\rm Im}\, \Gamma=-\sum_s{\rm tr}_p\, \ln(1-n_p) \quad ,
 \end{equation}
 where $n_p=\exp\left[-\frac{\pi}{e\, {\mathcal E}}(m^2+p_\perp^2)\right]$ is the mean number of pairs with momentum $p_\perp$ transverse to the direction of the field. The sum is over spin states $s$, and the momentum phase space trace becomes: ${\rm tr}_p\equiv  \frac{e{\mathcal E}L_\parallel T}{2\pi\hbar}V_\perp\int\frac{d^2p_\perp}{(2\pi\hbar)^2}$, in which we recognize the electric field analogue of the familiar Landau-level degeneracy factor. Typically, $n_p$ is very small, so we can approximate $-\ln(1-n_p)\approx n_p$, in which case we immediately obtain the leading Heisenberg-Euler result (\ref{constant-2}). On the other hand, expanding the logarithm we obtain the full instanton sum of Schwinger \cite{schwinger}
\begin{equation}
 \frac{{\rm  Im}\, \Gamma^{\rm HE}}{\rm Vol_4} \sim \hbar\, \frac{(e\,{\mathcal E})^2}{(2\pi)^3} \sum_{k=1}^\infty\frac{1}{k^2} \exp\left[-\frac{k\, \pi\, m^2}{e\, {\mathcal E}}\right] \quad .
 \label{instantonsum}
 \end{equation}
 This instanton sum expression can also be obtained by summing over all the poles at $s=n\,\pi$ in (\ref{hel}).
 
 This result is for spinor QED. If we were to consider the quantized charged particle to be a scalar particle, or even to have classical statistics, then the analogous expressions would be:
 \begin{equation}
 \frac{2}{\hbar}\, {\rm Im}\, \Gamma=
 \begin{cases} 
 {-\sum_s{\rm tr}_p\, \ln(1-n_p)\quad,\quad  {\rm Fermi}\cr\cr
{\rm tr}_p\, \ln(1+n_p)\quad,\quad {\rm Bose}\cr\cr
{\rm tr}_p\, n_p\qquad,\qquad  {\rm Boltzmann}}
 \end{cases}
 \end{equation}
Thus, when $n_p$ is small, the leading term is independent of statistics [apart from a  simple degeneracy factor $(2s+1)$], and so many computations work with scalar QED rather than spinor QED, since it is notationally simpler. But we should remember that $ \frac{2}{\hbar}\, {\rm Im}\,\Gamma \approx {\rm tr}_p\, n_p$ is an approximation. The higher order terms in the sum \eqn{instantonsum} correspond to the coherent production of multiple pairs \cite{nikishov-review,ritus-review,lebedev} in the same spacetime volume $\lambda_c^4$, and quantum statistics effects become important. It would be very interesting if these higher order correlations could be probed experimentally, but at present we are concentrating on being able to measure directly the first term.

\section{Beyond the constant field approximation: one-dimensional inhomogeneities}

\begin{figure}[h]
\centerline{\resizebox{0.6\textwidth}{!}{%
\includegraphics{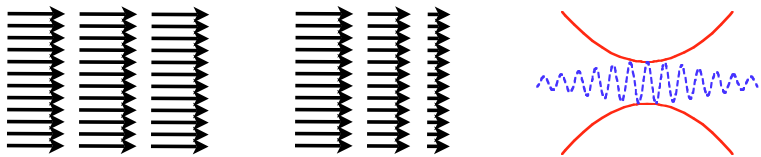}}
}
\caption{Progression from Heisenberg and Euler's original configuration of a uniform constant electric field, to more realistic  configurations with one-dimensional inhomogeneities, and eventually to realistic focussed laser pulses. Here, non-perturbative aspects of vacuum polarization are probed by the externally imposed electromagnetic  fields.}
\label{fig2}
\end{figure}

\begin{figure}[h]
\centerline{\includegraphics[scale=0.5]{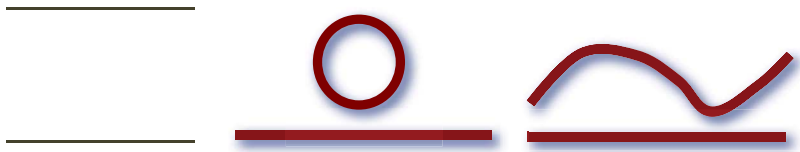}}
\caption{Progression from Casimir's original configuration of perfect parallel plates, to realistic experimental configurations with different geometries and imperfect surfaces. Here vacuum polarization is probed by the externally imposed boundaries.}
\label{fig3}
\end{figure}

So far, we have discussed the theoretical computation of the vacuum pair production probability only in the approximation of a constant and uniform applied electric field. But to reach the nonperturbative regime of ${\mathcal E}\sim {\mathcal E}_c$, for example  with ELI, will involve intense, short-pulse, focussed laser beams. Therefore, we need to understand how  the Heisenberg-Euler result (\ref{constant-2}) for ${\rm Im}(\Gamma[A])$ is modified for gauge fields $A_\mu(x)$ corresponding to realistic laser fields with strong spatial and temporal inhomogeneities, as depicted in Figure \ref{fig2}. This is a well-posed problem, for which the required formalism exists \cite{schwinger-qed,schwinger,salam}, in terms of determinants of the Dirac operator. It is nevertheless a very difficult computational problem for which there are few quantitative results. A somewhat analogous situation exists in the theory of the Casimir effect, as sketched in Figure \ref{fig3}. Casimir's original computation \cite{casimir} for a configuration of two parallel plane perfect mirrors leads to a simple expression for the Casimir energy and force, but for more realistic experimentally realizable surface configurations the computation is well-defined and easily posed, but it is extremely difficult  to obtain explicit quantitative results \cite{bordag}. 

\subsection{The Keldysh approach}

As a first step in this direction of computing the vacuum pair production probability in inhomogeneous external fields, recall once again the analogy between vacuum pair production and atomic ionization. In the context of atomic ionization, Keldysh \cite{keldysh} considered a monochromatic time dependent  electric field represented by the sinusoidal function ${\mathcal E}(t)={\mathcal E}\cos(\omega t)$. For an atom with binding energy $E_b$, the peak electric field amplitude ${\mathcal E}$ defines a frequency scale $\omega_K={e{\mathcal E}}/{\sqrt{2m E_b}}$, and Keldysh identified two important physical regimes characterized by the "Keldysh adiabaticity parameter"
\begin{equation}
\gamma_K \equiv \frac{\omega}{\omega_K}=\frac{\omega \sqrt{2m E_b}}{e{\mathcal E}} \quad .
\end{equation}
The region $\gamma_K \ll 1$ is an instantaneous, tunneling, nonperturbative regime, while the opposite limit where  $\gamma_K \gg 1$ turns out to be a perturbative multi-photon regime. Note that both regions, small and large $\gamma_K$, can be spanned without violating the relevant approximations:
\begin{eqnarray}
{\mathcal E}&\ll &{\mathcal E}_c^{\rm ionization}\equiv \frac{\sqrt{2m}E_b^{3/2}}{e\hbar}\quad [{\rm weak\,\, field\,\, approximation}]\nonumber\\
\hbar \omega&\ll &E_b \qquad\qquad [{\rm classical\,\, field\,\, approximation}]
\label{limits}
\end{eqnarray}
In the sinusoidally time dependent electric field, ${\mathcal E}(t)={\mathcal E}\cos(\omega t)$, Keldysh found the following simple approximate expression for the ionization probability:
\begin{eqnarray}
P_{\rm ionization}\sim\exp\left[-\frac{4}{3}\frac{\sqrt{2m}\,
E_b^{3/2}}{e{ \mathcal E}\hbar}\,g(\gamma_K^2)\right] 
 \label{kion}
 \end{eqnarray}
The function $g(\gamma_K^2)\sim 1$ for  $\gamma_K\ll 1$, while $g(\gamma_K^2)\sim 3 \ln (2 \gamma_K)/(2\gamma_K)$ for $\gamma_K\gg 1$. Thus, we obtain the limiting behaviors:
\begin{eqnarray}
P_{\rm ion.}\sim
\begin{cases}
{ \exp\left[-\frac{4}{3}\frac{\sqrt{2m}\,
E_b^{3/2}}{e{ \mathcal
E}\hbar}\right]
\,\, , \,\,
\gamma_K \ll 1 \,\,\, ({\rm nonperturbative})
\cr
\left(\frac{e{ \mathcal E}}{2\omega \sqrt{2m
E_b}}\right)^{2E_b/\hbar \omega}
\,\, , \,\,\gamma_K \gg 1 \,\, ({\rm perturbative})}
 \end{cases}
 \nonumber
 \label{kion-limits}
 \end{eqnarray}
 For small $\gamma_K$ the field ${\mathcal E}(t)$ is essentially constant over the time scale $1/\omega_K$, and the ionization process is a nonperturbative tunneling effect, as in Oppenheimer's result (\ref{oppie}). On the other hand, for large $\gamma_K$ it is a perturbative multi-photon effect, which can be seen from the fact that in this case the probability is given by the normalized perturbative field ${\mathcal E}$ raised to a power equal to (twice) the number of photons required to reach the binding energy.
Keldysh's work  was developed further by several authors \cite{perelomov,kotova,perelomov-book,faisal}, and the resulting theory to describe strong-field ionization in atomic and molecular systems is commonly referred to as the "ADK" theory \cite{adk} (see also \cite{popov-comment}).  However, an important comment should be kept in mind: in ultra-intense lasers it is not necessarily realistic to neglect the magnetic component of the laser field \cite{reiss}, so that the representation of the laser field by just a time-dependent electric field ${\mathcal E}(t)$ should be extended. We return to this comment below, in Section \ref{beyond}.

The Keldysh approach has been adapted to the QED vacuum pair production problem \cite{brezin,popov-wkb,popov-oscillator}, once again with the laser field modeled by  an oscillating electric field ${\mathcal E}(t)={\mathcal E}\cos(\omega t)$, and with the role of the binding energy $E_b$  played by $2mc^2$. The frequency scale set by the peak electric field amplitude is $\omega_{QED}={e{\mathcal E}}/{(mc)}$, and the analogous "adiabaticity parameter" is
\begin{equation}
\gamma \equiv \frac{\omega}{\omega_{QED}}=\frac{m c \omega}{e{\mathcal E}}\equiv \frac{1}{a_0} \quad .
\end{equation}
Here we note that we can also express the adiabaticity parameter $\gamma$ as the inverse of the "normalized field strength" parameter
\begin{equation}
a_0\equiv \frac{e{\mathcal E}}{m c \omega}\qquad .
\label{a0}
\end{equation}
The parameter $a_0$ provides a familiar characterization of the laser intensity in the laser and plasma physics literature. Using WKB techniques, one finds \cite{brezin,popov-wkb,popov-oscillator}, analogous to \eqn{kion},
\begin{equation}
P_{\rm pair \, prod.}\sim
\begin{cases}
 {\exp\left[-\frac{\pi m^2 c^3}{e{\mathcal E}\hbar}\right]
\,\, , \,\,
\gamma\ll 1 \,\, ({\rm nonperturbative})\cr
\left(\frac{e{\mathcal E}}{m\omega
}\right)^{2mc^2/\hbar \omega}\,\, , \,\,\gamma \gg 1 \,\, ({\rm perturbative})}
 \end{cases}
 \label{wkb}
 \end{equation}
 In the perturbative multi-photon regime, this QED pair production effect has been observed in a beautiful experiment (E-144) at SLAC \cite{burke}, in which a laser pulse collided with the (highly relativistic) SLAC electron beam, leading to nonlinear Compton scattering involving 5 photons, producing a high energy gamma photon that decays into an electron-positron pair. An important challenge for the ELI project will be to observe  pair  production {\it in the nonperturbative regime}, and {\it directly from vacuum}, as a projected goal of ELI \cite{eli} is to attain an adiabaticity parameter $\gamma\sim 1/5000\ll 1$. Theoretically, this requires reliable predictions of pair production rates and spectra in this nonperturbative regime, for specific ELI laser parameters. Before discussing new work in this direction I review various approaches that have been developed to analyze the case of an external field that is inhomogeneous in one direction.
 
 \subsection{The one-dimensional scattering picture}
 
 
 An important perspective on vacuum pair production in electric fields with a one-dimensional inhomogeneity is obtained by viewing the problem as  a one dimensional quantum mechanical scattering problem (Popov \cite{popov-oscillator} and Perelomov and Zel'dovich \cite{perelomov-book} each attribute this idea in this context  to L.~Pitaevskii). This QM scattering problem can then be solved exactly in some cases, numerically in most cases, or semiclassically by various WKB approaches. Consider a time dependent electric field represented by the vector potential $A_3(t)$, so that ${\mathcal E}_3(t)=-\dot{A}_3(t)$.
 As mentioned above, for the leading contribution to the pair production probability we can simplify matters by working with scalar QED, so we write the Klein-Gordon equation $(-D_\mu D^\mu +m^2)\phi=0$ for the field operator $\phi$ as
 \begin{equation}
 -\ddot{\phi}-(p_3-eA_3(t))^2\phi=(m^2+p_\perp^2)\phi
 \label{pit}
\end{equation}
This can be viewed as a one dimensional quantum mechanical Schr\"odinger equation, with $t$ playing the usual role of $x$.  To describe  the process of pair production, we seek solutions of (\ref{pit}) with just positive frequencies in the far future, but a mixture of positive and negative frequencies as $t\to -\infty$:
\begin{figure}[h]
\centerline{\resizebox{0.4\textwidth}{!}{%
\includegraphics{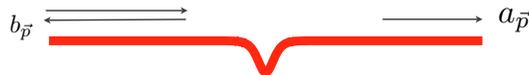}}
}
\caption{The Bogoliubov transformation between in and out states at $t=\pm \infty$, with boundary conditions in (\ref{bcs}), can be viewed as a one dimensional quantum mechanical scattering problem.}
\label{fig4}
\end{figure}
 \begin{eqnarray}
 \phi&\sim & e^{-it\sqrt{m^2+p^2}}+b_{\vec{p}}\,
e^{it\sqrt{m^2+p^2}}\quad , \quad t\to-\infty
\nonumber\\
&\sim & a_{\vec{p}}\, e^{-it\sqrt{m^2+p^2}}\quad , \hskip 2cm
t\to+\infty
\label{bcs}
\end{eqnarray}
In the QM picture, this looks like a scattering problem, 
and the Bogoliubov transformation between $t=\pm\infty$ modes is encoded in the reflection amplitude $b_{\vec{p}}$. Since the "potential", $-(p_3-e A_3(t))^2$, is negative, and the "energy", $(m^2+p_\perp^2)$, is positive, this is the case of over-the-barrier scattering \cite{pokrovskii}, and so the reflection coefficient $|b_{\vec{p}}|^2$ is expected to be exponentially small. By familiar manipulations \cite{brezin,popov-wkb,popov-oscillator,hagen}, the pair production probability is given by 
\begin{eqnarray}
P&\approx &\int \frac{d^3 p}{(2\pi)^3}\, |b_{\vec{p}}|^2\\
&\approx & \int \frac{d^3 p}{(2\pi)^3}\, e^{-2\,{ \rm
Im}\oint\sqrt{m^2+p_\perp^2+[p_3-e A_3(t)]^2}\,\,dt } \nonumber
\end{eqnarray}
The dominant exponential contribution can be extracted by setting $\vec{p}=0$, while the prefactor comes from Gaussian integration in the momenta, after expanding the integral in the exponent to quadratic order in momenta. The transverse and longitudinal momenta contribute differently, and one arrives at a simple compact formula  \cite{popov-wkb}:
\begin{eqnarray}
{\rm Im}\,\Gamma^{\rm WKB}
\approx V_3 \, \frac{\sqrt{2}(e{\mathcal E})^{5/2}}{32\pi^3m\omega}\, \frac{\exp\left[-\frac{m^2\pi}{e {\mathcal E}}\,g(\gamma^2)\right]}{\frac{d}{d(\gamma^2)} \left(\gamma^2\, g(\gamma^2)\right)\, \sqrt{-\frac{d^2}{d(\gamma^2)^2} \left(\gamma^2\, g(\gamma^2)\right)}}
\label{wkb-1}
\end{eqnarray}
Here the function $g(\gamma^2)$ is analogous to the function $g(\gamma_K)$ appearing in Keldysh's expression (\ref{kion}). An explicit formula for $g$ is:
\begin{equation}
g(\gamma^2)\equiv \frac{2}{\pi} \int_{-1}^1 dy\,\frac{\sqrt{1-y^2}}{|f^\prime |}
\end{equation}
where we have written the imaginary time version of the gauge potential as $\frac{{\mathcal E}}{\omega}\, f(i t)$, defined $y=f/\gamma$, and it is understood that $f^\prime$ is written back in terms of $y$. For example, for the cosine field ${\mathcal E}(t)={\mathcal E}\cos(\omega t)$, we have $A_3(t)=-\frac{{\mathcal E}}{\omega}\sin(\omega t)$, and so $f(x_4)=\sinh(\omega x_4)$. Thus, with $y=f/\gamma$, we have  $f^\prime=\cosh(\omega x_4)=\sqrt{1+\gamma^2 y^2}$, and so 
\begin{eqnarray}
g(\gamma^2)&=& \frac{4\sqrt{1+\gamma^2}}{\pi \gamma^2}\left[{\bf K}\left(\frac{\gamma^2}{1+\gamma^2}\right)-{\bf E}\left(\frac{\gamma^2}{1+\gamma^2}\right)\right]\nonumber\\
&\sim &
\begin{cases}
{1-\frac{1}{8}\gamma^2 \quad, \quad \gamma\ll 1\cr
\frac{4}{\pi \gamma}\ln \gamma \quad , \quad \gamma\gg 1
}
\end{cases}
\label{wkb-2}
\end{eqnarray}
Here ${\bf K}$ and ${\bf E}$ are the standard complete elliptic integrals. The result in (\ref{wkb-2}) explains the tunneling and multiphoton limits quoted in (\ref{wkb}).
Notice that in the WKB formula (\ref{wkb-1}), the exponent and the prefactor are both expressed in terms of the same function $g(\gamma^2)$. This result is typically obtained in the imaginary time method of WKB \cite{popov-wkb,popov-oscillator}, but can also be found in conventional WKB \cite{brezin}, and in the phase integral method \cite{kimpage}, where higher order WKB contributions have also been considered. A similar formalism exists for static electric fields that depend on one spatial coordinate, represented for example by the gauge field $A_0(x_3)$ \cite{nikishov-spatial,kimpage,hagen}. It is useful to compare these semiclassical results with some exact results \cite{naro,nikishov-spatial} that arise for the cases ${\mathcal E}(t)={\mathcal E}\,{\rm sech}^2(\omega t)$ or ${\mathcal E}(x_3)={\mathcal E}\,{\rm sech}^2(k\,x_3)$.  For these fields, the associated Klein-Gordon equation is a soluble Schr\"odinger problem and so the effective action can be computed in closed form. (In fact, the associated Dirac equations are also exactly soluble). The semiclassical results compare very well with these exact results in the  semiclassical regime \cite{dunnehall,kimpage}. For example, using the phase integral method, for the soluble cases ${\mathcal E}(t)={\mathcal E}\,{\rm sech}^2(\omega t)$ or ${\mathcal E}(x_3)={\mathcal E}\,{\rm sech}^2(k\,x_3)$, respectively, the particle number can be expressed as \cite{kimpage}
\begin{eqnarray}
{\mathcal N}
&\approx&
\frac{(e {\mathcal E})^{5/2} T}{4\pi^3 m}
\left(1+\gamma^2 \right)^{5/4}
\exp\left[
-\frac{\pi m^2}{e {\mathcal E}}
\left(\frac{\sqrt{1+\gamma^2}-1}
{2\gamma^2}
\right)
\right] 
\nonumber\\
{\mathcal N}
&\approx&
\frac{(e {\mathcal E})^{5/2} L}{4\pi^3 m}
\left(1-\tilde{\gamma}^2 \right)^{5/4}
\exp\left[
-\frac{\pi m^2}{e {\mathcal E}}
\left(\frac{1-\sqrt{1-\tilde{\gamma}^2}}
{2\tilde{\gamma}^2}
\right)
\right]
\nonumber
\end{eqnarray}
where in the first expression $\gamma=\frac{m \omega}{e{\mathcal E}}$, while in the second expression $\tilde{\gamma}=\frac{m k}{e{\mathcal E}}$. 
When the gamma factors go to 0 we recover the form of the constant field Heisenberg-Euler result, as is explained in detail in \cite{kimpage}.
These results illustrate an important point: a temporal inhomogeneity tends to enhance  particle production, while a spatial inhomogeneity tends to suppress particle production. This follows from the different behaviors in the above formulas of the exponents and of the prefactors. Physically, this can be understood as follows:  with a {\it spatial} electric field, the dominant nucleation of pair production occurs at a maximum of the field, but the virtual particles are then accelerated apart, away from this point, and if the field falls off too quickly then they may not gain enough energy to become real particles, so that the pair production is suppressed. This effect is seen clearly in the vanishing of the particle number when $\tilde{\gamma}\to 1$. On the other hand, with a time-dependent field we can regard this as a tunneling problem with an oscillating barrier, for which the tunneling is enhanced roughly speaking because the average barrier is lower \cite{popov-oscillator,perelomov-book}.

\subsection{Quantum Kinetic Approach}

Another widely-studied approach in the analysis of the  time dependent electric field case, ${\mathcal E}={\mathcal E}(t)$, is the quantum kinetic approach \cite{kluger,schmidt,bloch,alkofer,florian-1}. 
The key quantity in this formalism is the momentum distribution function, ${\mathcal N}(\vec{p}, t)$, of pairs at a time $t$, which satisfies a quantum Vlasov equation (here for spinor QED):
\begin{eqnarray}
\frac{d {\mathcal N}}{dt}&=&\frac{e^2}{2}(m^2+p_\perp^2)\frac{{\mathcal E}(t)}{\omega^2(t)}\int_{-\infty}^t dt^\prime\, \frac{{\mathcal E}(t^\prime)}{\omega^2(t^\prime)}\left(1-2{\mathcal N}(t^\prime)\right)  \cos\left(2\int_{t^\prime}^t d\bar{t}\, \omega(\bar{t})\right)
\label{vlasov}
\end{eqnarray}
Here we have defined
\begin{equation}
\omega^2(t)=m^2+p_\perp^2+(p_3 - e A_3(t))^2
\label{omega}
\end{equation}
For scalar QED the analogous quantum Vlasov equation is:
\begin{eqnarray}
\frac{d {\mathcal N}_{\rm sc}}{dt}&=&\frac{1}{2}\frac{\dot{\omega}(t)}{\omega(t)}\int_{-\infty}^t dt^\prime\, \frac{\dot{\omega}(t^\prime)}{\omega(t^\prime)}\left(1+2{\mathcal N}_{\rm sc}(t^\prime)\right)  \cos\left(2\int_{t^\prime}^t d\bar{t}\, \omega(\bar{t})\right)
\label{vlasov-sc}
\end{eqnarray}
The asymptotic large-time value of ${\mathcal N}(\vec{p}, t)$ gives the momentum distribution of the produced pairs. Note that since the field is spatially uniform we can label the produced pairs by their momentum. The total particle number is obtained by integrating ${\mathcal N}(\vec{p}, \infty)$  over the momenta. The form of these quantum kinetic equations shows that the process is inherently non-Markovian and non-local in time \cite{kluger}. Furthermore, back-reaction effects have been studied in this approach \cite{bloch}, showing that they are negligible if the maximum  amplitude of the electric field is lower than the critical field ${\mathcal E}_c$ \cite{florian-thesis}. 

It is not as widely appreciated that the quantum kinetic approach is equivalent to the Popov/Pitaevskii one dimensional scattering approach described in the previous section \cite{dumlu}. We illustrate this briefly here for scalar QED. Recall that we can write the Klein-Gordon equation (\ref{pit}) as
\begin{eqnarray}
\ddot{\phi}+\omega^2(t) \phi=0
\label{pit-2}
\end{eqnarray}
and convert this Schr\"odinger-like scattering problem into a Riccati equation as follows  \cite{popov-oscillator}. First, express the complex field $\phi$ as
\begin{eqnarray}
\phi\equiv A\left(e^{-i\varphi}+R e^{i\varphi}\right) \quad ; \quad \varphi(t) \equiv \int_{-\infty}^t \omega(t^\prime) d t^\prime
\label{popov-phi}
\end{eqnarray}
in terms of two other complex fields $R$ and $A$ where
\begin{eqnarray}
\frac{\dot{A}}{A}=-\frac{\dot{R}e^{i\varphi}}{e^{-i\varphi}+R\,e^{i\varphi}} \quad .
\label{ar}
\end{eqnarray}
Then 
\begin{equation}
\dot{\phi}=-i\omega A\left(e^{-i\varphi}-r\,e^{i\varphi}\right)\quad ,
\label{phidot}
\end{equation}
and the scattering equation (\ref{pit}) becomes a Riccati equation for the scattering amplitude $R$:
\begin{eqnarray}
\dot{R}=\frac {\dot{\omega}}{2\omega}\left(e^{-2i\varphi}-R^2 e^{2i\varphi}\right) \quad .
\label{popov-r}
\end{eqnarray}
It is straightforward to integrate this equation numerically, with initial condition $R(-\infty)=0$, and obtain the probability for particle production as $| R(\infty)|^2$. Since (\ref{pit}) has the form of a Schr\"odinger scattering problem with an inverted potential, this is the case of over-the-barrier scattering, which is conveniently analyzed with the method of imaginary time \cite{popov-imag}. On the other hand, it could just as well be solved "exactly", by numerical integration.

The connection with the quantum kinetic approach is seen  by noting that in the quantum kinetic approach [for scalar QED], one writes the field $\phi$ in terms of two other complex fields $\alpha$ and $\beta$:
\begin{eqnarray}
\phi&\equiv& \frac{\alpha}{\sqrt{2\omega}}e^{-i\varphi}+ \frac{\beta}{\sqrt{2\omega}}e^{i\varphi}\nonumber\\
\dot{\phi}&=& -i\omega\left(\frac{\alpha}{\sqrt{2\omega}}e^{-i\varphi}- \frac{\beta}{\sqrt{2\omega}}e^{i\varphi} \right) \quad .
\label{qk-phi}
\end{eqnarray}
The second equation requires $\alpha$ and $\beta$ to be related by:
\begin{eqnarray}
\dot{\alpha}=\frac{\dot{\omega}}{2\omega} \, \beta \, e^{2i\varphi}\quad , \quad 
\dot{\beta}=\frac{\dot{\omega}}{2\omega} \, \alpha \, e^{-2i\varphi} \quad .
\label{alpha-beta}
\end{eqnarray}
Comparing the two decompositions for $\phi$ in (\ref{popov-phi}) and (\ref{qk-phi}) we see that $R\equiv \beta/\alpha$, and the equations for $\alpha$ and $\beta$ in (\ref{alpha-beta}) are then equivalent to the Riccati equation (\ref{popov-r}). The particle number is
\begin{eqnarray}
{\mathcal N}_{\rm sc}=|\beta |^2=\frac{|R|^2}{1-|R|^2} \quad ,
\label{n-r}
\end{eqnarray}
using the scalar relation: $|\alpha|^2-|\beta|^2=1$. Converting the Riccati equation for the scattering amplitude $R$ into an equation just for the probability $|R|^2$,  one finds that it is in fact completely equivalent to the quantum Vlasov equation (\ref{vlasov-sc}) of scalar QED \cite{dumlu}. Thus, the quantum kinetic approach may be viewed as  a real time version of Popov's imaginary time scattering problem. This can also be described equivalently in terms of Green's functions and the S-matrix \cite{gavrilov}, or in a Schr\"odinger picture \cite{padman}.

\subsection{Pulse shaping}

In going beyond the constant field approximation, the goal is to understand how inhomogeneities modify the constant field result (\ref{constant}), so that one might design a pulse shape that leads to a higher yield of vacuum pair production, at a lower peak field intensity. The importance of pulse shape for these processes has been studied widely \cite{popov-shape,dipiazza}, and reflects experience gained in the strong-field atomic and molecular physics community, where sophisticated pulse-engineering techniques have been developed for ionization processes \cite{becker,gerber}.

In several recent papers \cite{bulanov}, estimates based on taking time dependent laser pulse(s), neglecting the spatial variation in the focus or the interaction region, suggest that the intensity at which one could see the nonperturbative effects of vacuum pair production could be a factor of 100 smaller than the critical intensity $I_c\approx 4\times 10^{29}\,{\rm W/cm^2}$. This is once again reminiscent of the situation for atomic ionization, where ionization effects can be observed at intensities lower than the critical field estimate of $10^{16}\,{\rm W/cm}^2$ coming from Oppenheimer's constant field result (\ref{oppie}). In the QED vacuum polarization problem, with focussed laser pulses, the field polarization plays an important role \cite{bulanov}, but the essential enhancement mechanism is that the interaction region is very large compared to the scale of the Compton wavelength, and so there is a significant  prefactor enhancement. In these papers, the authors work with an external field that is an exact solution of Maxwell's equations, and that mimics a Gaussian laser beam; they then convolve this solution with a pulse time-envelope function to obtain an approximate solution that represents a focussed pulsed laser beam. In the vicinity of the focus spot, they argue that the spatial dependence can be neglected, so that in the end the problem  becomes essentially one-dimensional, and the semiclassical results can be used. An important feature of this approach is that the role of the polarization of the field is clearly incorporated, and the cases of a single focussed beam, and of two counter-propagating beams were analyzed. The case of two  counter-propagating circularly polarized beams leads to a higher yield of produced pairs, and the threshold intensity at which one might expect to see such effects is several orders of magnitude lower than the Schwinger limit. This is encouraging news for ELI. 

In a related development, it has recently been shown that the superposition of two time-dependent electric fields can lead to a significant enhancement of vacuum pair production, a "dynamically assisted Schwinger mechanism'' \cite{dyn}. The specific example studied in \cite{dyn} consists of a combination of  a strong but slow field pulse,  and a weak but fast field pulse: 
\begin{eqnarray}
{\mathcal E}_{\rm slow}(t)&=&{\mathcal E}\,{\rm sech}^2(\Omega t)\nonumber\\
{\mathcal E}_{\rm fast}(t)&=&\epsilon\, {\rm sech}^2(\omega t)\nonumber
\label{dyn-fields}
\end{eqnarray}
The parameters fall into the hierarchies:
\begin{eqnarray}
&&0<\epsilon \ll {\mathcal E}\ll {\mathcal E}_c\nonumber\\
&&0<\Omega \ll \omega \ll m \nonumber
\label{hierarchy}
\end{eqnarray}
Surprisingly, even though both field amplitudes, ${\mathcal E}$ and $\epsilon$, are below the critical field ${\mathcal E}_c$ in (\ref{critical}), there is significant enhancement of the pair production rate when the frequencies are also related as in (\ref{hierarchy}). The non-perturbative pair production process that we would associate with the slow strong field interacts with the perturbative multiphoton pair production process that we would associate with the fast weaker field to produce a stronger impact than each process separately. This illustrates another possible approach to shaping the laser pulses in the time domain so as to enhance the pair production yield.

\section{Towards more general laser fields}
\label{beyond}

From a theoretical point of view, as well as from a practical experimental point of view, we should ask if we can go beyond the restriction to a one dimensional field inhomogeneity discussed in the previous section. There are several urgent reasons for studying more general background fields. First, such fields are needed to model more realistically the intense, focussed, short-pulse laser fields to be produced at ELI. Second, in the ultra-relativistic regime of ELI it is by no means clear that one can continue to neglect the magnetic component of the laser field and consider just the electric component. These questions need to be answered.

In fact, surprisingly little is known about vacuum pair production in external fields that have a more complicated inhomogeneity structure, such as appears in a more realistic laser field. The results of all the QED work summarized in the previous section is that when the background field is approximated as an electric field pointing in one fixed direction, but with amplitude that either (i) varies with time, or [but not both] (ii) varies with space only along the direction of the field, then the problem can be reduced to a one-dimensional problem that can solved numerically, or approximately with WKB. However,  all these standard approaches become practically intractable when the field depends on more than one coordinate. An interesting exception is for fields that depend on a light-cone coordinate $(x- c t)$ \cite{tomaras,fried}.

\subsection{Worldline formalism for the effective action}

It is  possible to develop semiclassical approximations for the case of multi-dimensional inhomogeneities in the external field, using Feynman's worldline formalism of the effective action. Feynman \cite{feynman} formulated a first-quantized form of QED,  which amounts to representing \cite{strassler,csreview} the effective action as a quantum mechanical path integral over closed loops $x_\mu(\tau)$ in four dimensional spacetime, with the closed loops being parametrized by the proper time $\tau$. As Feynman noted, this proper-time representation is the natural representation for maintaining relativistic invariance. The propertime parametrization had  been developed earlier by Fock and Nambu \cite{fock}, and was  also used by Schwinger, in operator form rather than in path integral form,  in his landmark QED computation of vacuum pair production \cite{schwinger}.
Feynman's  worldline path integral formalism has since been extended significantly, primarily for applications in {\it perturbative} quantum field theory \cite{csreview}, building on analogies and motivation from the Polyakov formulation of string theory. This rebirth has led to many beautiful advances in our understanding of perturbative scattering amplitudes \cite{bern}, but here I propose to use it instead to extract {\it non-perturbative} information. Let us consider scalar QED rather than spinor QED, as it is notationally simpler, and the leading imaginary part of the effective action only differs from that of spinor QED by a spin degeneracy factor of 2. Then the effective action for a scalar charged particle (charge $e$, mass $m$) in a Euclidean classical gauge background $A_\mu(x)$ is the functional: 
\begin{eqnarray}
\Gamma [A] &=&-{\rm tr} \, \ln \left(-D_\mu^2+m^2\right)\nonumber\\
&=& \int_0^{\infty}\frac{dT}{T}\, e^{-m^2T}\int d^4x^{(0)} \, \langle x^{(0)}| e^{-T(-D_\mu^2)}\, |x^{(0)}\rangle \nonumber \\
&=&
\int_0^{\infty}\frac{dT}{T}\, e^{-m^2T}\int d^4x^{(0)} \hskip -.5 cm \int\limits_{x(T)=x(0)=x^{(0)}}\!\!\!\!\!\!\!\!\!\! {\mathcal D}x
\,\, {\rm exp}\left[-\int_0^Td\tau
\left(\frac{\dot x^2_\mu}{4} +i\, e\, A_\mu \dot x_\mu \right)\right]
\label{feynman}
\end{eqnarray}
To understand the origin of this expression, recall the standard representation of the logarithm of an operator as: $\ln {\mathcal M}=-\int_0^\infty \frac{dT}{T}\, e^{-T\, {\mathcal M}}$. This explains the second line in (\ref{feynman}), where we have performed the trace in the position representation. The last line in (\ref{feynman}) follows from interpreting the matrix element $\langle x^{(0)}| e^{-T(-D_\mu^2)}\, |x^{(0)}\rangle $ as a Euclidean transition amplitude, which is then written using Feynman's representation as a path integral, but now involving paths $x_\mu$  in four dimensional space. Since for the effective action we compute the {\it trace} of the logarithm, the matrix element is between identical points $x^{(0)}$, and so the path integral is over closed loops. Thus, we arrive at a first-quantized representation of the effective action $\Gamma[A]$
in terms of a functional integral $\int {\mathcal D}x$ over all closed Euclidean spacetime paths $x_\mu(\tau)$ that are periodic (with period $T$) in the proper-time parameter $\tau$ \cite{feynman}.  I use the QED worldline path integral normalization conventions of \cite{csreview}.  Note also that we have expressed the path integral in Euclidean space, as is appropriate for describing  a non-perturbative process.

This representation of the effective action translates the primary  computational problem into one of computing the quantum mechanical path integral in the last line of  (\ref{feynman}). This simple observation has several advantages. First, one could evaluate this Euclidean path integral numerically, using the worldline Monte Carlo technique of Gies and Langfeld \cite{gieslangfeld}. One creates a statistical ensemble of configurations, namely a set of closed loops $x_\mu(\tau)$ in 4 dimensional Euclidean spacetime, weighted by a factor $\exp[-\int^T_0 d\tau \dot{x}^2/4]$, and then one computes numerically, using a Monte Carlo algorithm,  the expectation of the "Wilson loop" operator $\exp[-e\int_0^Td\tau  A_\mu \dot{x}_\mu]$. For the pair production problem, the exponentially small imaginary part, arising from poles in the $T$ integral, can then be extracted numerically. This is still a challenging numerical problem, but it  has been shown to work well for situations with one-dimensional inhomogeneities of the electric field \cite{wl-pp}. This is potentially a very powerful approach, as it does not rely on any particular symmetry of the background field -- the ensemble of closed spacetime loops used in the numerical  Monte Carlo evaluation of the quantum mechanical path integral  is independent of the background field. Physically, the ensemble  probes the structure and impact of the background field. This means that in principle the numerical worldline Monte Carlo approach can be used for quite general inhomogeneous laser field configurations. This approach deserves further exploration.

\subsection{Worldline Instantons}

A second, complementary, approach is more analytical:  make a {\it semiclassical approximation} to the Euclidean worldline path integral. The essential idea is that the quantum mechanical path integral may be approximated by the contribution of  a small number [possibly just one] of  classical closed loops, plus the quantum fluctuations about these loops. These dominant Euclidean loops are called "worldline instantons". This is the Euclidean space analogue of the familiar semiclassical scattering theory \cite{pechukas,kleinert}, where a real-time scattering process is approximated by classical trajectories, with corrections from quantum fluctuations about the classical trajectories. The dominance of the path integral by just  a few closed paths is a semiclassical approximation, valid when the classical action is large compared to $\hbar$. If the background field has a simple functional form, depending on just a few physical parameters such as peak field amplitude, a frequency or wavenumber, etc..., then the regime of the semiclassical approximation can be precisely quantified in terms of a few dimensionless quantities.  The idea of seeking dominant semiclassical solutions to the path integral appearing in (\ref{feynman}) was suggested long ago \cite{halpern}, and was first applied to the vacuum pair production problem for a constant electric field in \cite{affleck}. More recently, the idea has been extended to the vacuum pair production problem with inhomogeneous background field configurations \cite{wli1,wli2,wli3,wli-review}. The special dominant loops are solutions to the associated classical (Euclidean) equations of motion for a charged particle moving in the given [inhomogeneous] electromagnetic field $F_{\mu\nu}(x)$\footnote{Note that the factor $2e$ has nothing to do with Cooper pairs; the factor of 2 is simply a side-effect of the standard normalization conventions of worldline path integrals \cite{csreview}, in which the "kinetic" term in (\ref{feynman}) is $\dot{x}^2/4$, rather than $\dot{x}^2/2$.}:
\begin{eqnarray}
\ddot{x}_\mu=2\,i\,e\, F_{\mu\nu}(x)\, \dot{x}_\nu \quad, \quad (\mu,\nu=1\dots4)\quad .
\label{euler}
\end{eqnarray}
Solutions of the classical Euclidean equations of motion \eqn{euler}
that are closed and periodic  are called {\it worldline instantons}. 

\begin{figure}[h]
\centerline{\resizebox{0.3\textwidth}{!}{%
\includegraphics{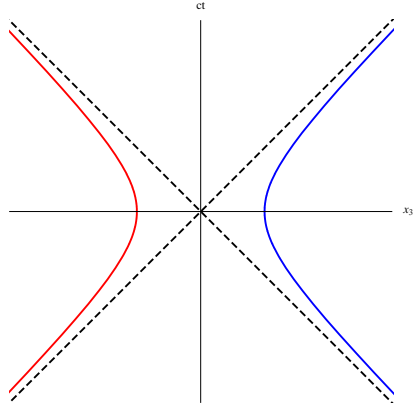}}
}
\caption{The Minkowski space-time trajectories of a charged particle in a constant electric field. The analogy with the trajectory of an accelerated observer is the starting point for the discussion of the Unruh effect in strong field laser systems -- see R. Sch\"utzhold's article in this Volume.}
\label{fig5}
\end{figure}

\begin{figure}[h]
\centerline{\resizebox{0.3\textwidth}{!}{%
\includegraphics{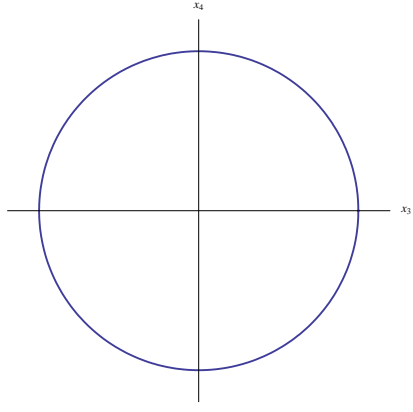}}
}
\caption{The Euclidean space-time trajectories [both directions around the loop] of a charged particle in a constant electric field. The open trajectories in Minkowski space, shown in Figure \ref{fig5}, are now closed trajectories in Euclidean space. This is the simplest worldline instanton, a periodic solution to the classical Euclidean equations of motion.}
\label{fig6}
\end{figure}

\begin{figure}[h]
\centerline{\resizebox{0.3\textwidth}{!}{%
\includegraphics{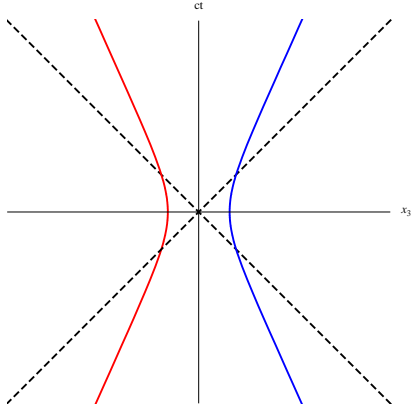}}
}
\caption{The Minkowski space-time trajectories [solid lines] of a charged particle in an electric field  pulse, ${\mathcal E}(t)={\mathcal E}{\rm sech}^2(\omega t)$, with adiabaticity parameter $\gamma=2$.}
\label{fig7}
\end{figure}

\begin{figure}[h]
\centerline{\resizebox{0.3\textwidth}{!}{%
\includegraphics{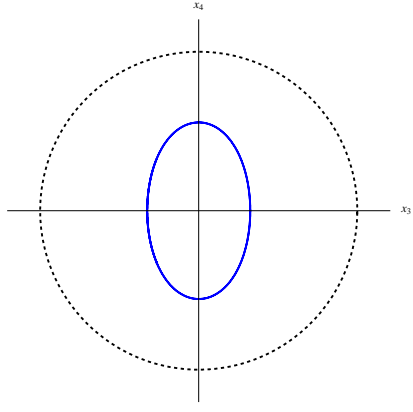}}
}
\caption{The Euclidean space-time trajectories [both directions around the loop] of a charged particle in an electric field  pulse,  ${\mathcal E}(t)={\mathcal E}{\rm sech}^2(\omega t)$, with adiabaticity parameter  $\gamma=2$. For reference, the dotted line shows the corresponding circular trajectory in a  constant field where $\gamma=0$. The open trajectories in Minkowski space are now closed trajectories in Euclidean space. This is the simplest worldline instanton, a periodic solution to the classical Euclidean equations of motion.}
\label{fig8}
\end{figure}

For example, for a constant (Minkowski) electric field, ${\mathcal E}_3={\mathcal E}$, the  Euler-Lagrange equations of motion in Minkowski space, with the coordinates $x_\mu(s)$ parametrized by proper time $s$ are (here the overdot $^.$ refers to $\frac{d}{ds}$):
\begin{eqnarray}
\ddot{x}_3=-2 e{\mathcal E}\, \dot{t}\quad , \quad \ddot{t}=-2\, e{\mathcal E}\,\dot{x}_3
\end{eqnarray}
The solution is well-known:
\begin{eqnarray}
x_3=\mp \frac{\cosh\left(2 e{\mathcal E}\, s \right)}{2e{\mathcal E}}\quad ,\quad
t=\pm \frac{\sinh\left(2 e{\mathcal E}\, s \right)}{2e{\mathcal E}}
\end{eqnarray}
In a $1+1$ dimensional Minkowski spacetime diagram, these follow open hyperbolic trajectories, as shown in Figure \ref{fig5}. On the other hand, in Euclidean space, with $x_4=i\,c\,t$, and $s\to i\tau$, the equations read (here the overdot $^.$ refers to $\frac{d}{d\tau}$):
\begin{eqnarray}
\ddot{x}_3=-2 e{\mathcal E}\,\dot{x}_4\quad , \quad \ddot{x}_4= 2e{\mathcal E}\,\dot{x}_3
\end{eqnarray}
and the solutions describe closed circular  orbits, as shown in Figure \ref{fig6}:
\begin{eqnarray}
x_3=\pm \frac{\cos\left(2 e{\mathcal E}\, \tau \right)}{2e{\mathcal E}}\quad ,\quad
x_4=\pm \frac{\sin\left(2 e{\mathcal E}\, \tau \right)}{2e{\mathcal E}}
\end{eqnarray}
For a  time dependent (Minkowski) pulse field, given by ${\mathcal E}_3(t)={\mathcal E}\,{\rm sech}^2(\omega t)$, the Minkowski equations are:
\begin{eqnarray}
\ddot{x}_3&=&-2 e {\mathcal E}\, {\rm sech}^2(\omega \, t)\,\dot{t}\nonumber\\
\ddot{t}&=&-2 e {\mathcal E}\, {\rm sech}^2(\omega \, t)\,\dot{x}_3
\end{eqnarray}
The Minkowski
solutions are the open trajectories:
\begin{eqnarray}
\hskip -.5cm x_3(s)&=&\mp \frac{1}{\omega \sqrt{1+\gamma^2}} \, {\rm arcsinh}\left[\gamma\, \cosh\left(2 e {\mathcal E} \sqrt{1+\gamma^2}\,s \right)\right]
\nonumber\\
\hskip -.75cm t(s)&=&\pm \frac{1}{\omega}\,  {\rm arcsinh}\left[\frac{\gamma}{\sqrt{1+\gamma^2}}\, \sinh\left(2 e {\mathcal E} \sqrt{1+\gamma^2}\, s \right)\right]
\end{eqnarray}
where $\gamma=\frac{m\omega}{e{\mathcal E}}$. These trajectories are shown in Figure \ref{fig7}. On the other hand, 
 the Euclidean trajectories are closed loops:
\begin{eqnarray}
x_3(\tau)&=&\frac{1}{\omega}\,\frac{1}{\sqrt{1+\gamma^2}} \, {\rm arcsinh}\left[\gamma\, \cos\left(2 e {\mathcal E} \sqrt{1+\gamma^2}\,\tau\right)\right]
\nonumber\\
x_4(\tau)&=&\frac{1}{\omega}\,  \arcsin\left[\frac{\gamma}{\sqrt{1+\gamma^2}}\, \sin\left(2 e {\mathcal E} \sqrt{1+\gamma^2}\, \tau\right)\right]
\end{eqnarray}
and are shown in Figure \ref{fig8}.

Expanding in the fluctuations about the worldline instanton loop, $x_{\mu}(\tau)=x_{\mu}^{\rm cl}(\tau) + \eta_{\mu}(\tau)$, we obtain an approximate expression for the path integral:
\begin{eqnarray}
\int {\mathcal D}x 
\,\, e^{-S[x]}
&\approx&   e^{-S[x^{\rm cl}]} \int{\mathcal D}\eta 
\,\, \exp\left[-\frac{1}{2} \int_0^T d\tau\, \eta_\mu \Lambda_{\mu\nu} \eta_\nu\right]
\nonumber\\
&=& \frac{e^{-S[x^{\rm cl}]}}{\sqrt{{\rm Det}\, \Lambda}}\quad ,
\label{fluc}
\end{eqnarray}
where $S[x^{\rm cl}]$ is the classical action, evaluated on the worldline instanton loop.
The prefactor is expressed in terms of the determinant of the  fluctuation operator (the "Jacobi operator") is 
\begin{equation}
\Lambda_{\mu\nu}\equiv -\frac{1}{2}\,\delta_{\mu\nu}\, \frac{d^2}{d\tau^2}+ e\, F_{\mu\nu}(x^{\rm cl})\, \frac{d}{d\tau}+e\, \partial_\mu F_{\rho\nu}(x^{\rm cl})\, \dot{x}^{\rm cl}_\rho\quad .
\label{morse}
\end{equation}
In \cite{wli1} it was shown that the worldline instanton exponential factor $e^{-S[x^{\rm cl}]}$ agrees precisely with the WKB exponential factor $\exp[-\frac{m^2 \pi}{e{\mathcal E}}\,g(\gamma^2)]$ in (\ref{wkb-1}), for the case of one-dimensional inhomogeneities. 

The form of the fluctuation operator in (\ref{morse})  exhibits another major advantage of the worldline approach -- since the paths are parametrized by propertime, the fluctuation problem refers to a one-dimensional operator (actually, coupled one-dimensional operators) rather than a partial differential operator in four dimensional spacetime. This is technically important because there are simple mathematical methods for computing the determinant of such one-dimensional operators \cite{gelfand,levit,kirsten,kleinert}, but these techniques do not generalize easily to partial differential operators. Thus one can numerically compute ${\rm det}(\Lambda)$ for any period $T$, and then perform the remaining  $T$ and $x^{(0)}$ integrals in the effective action expression \eqn{feynman}. In fact, in \cite{wli2} it was shown that for one-dimensional inhomogeneities the determinant of the fluctuation operator can be computed analytically:
\begin{eqnarray}
{\rm Det}(\Lambda)= \left( 2\, m\, 
\dot{x}_4^{\rm cl}(T) \int_0^T\frac{d\tau}{\left[\dot{x}_4^{\rm cl}(\tau)\right]^2}\right)^2
\label{22}
\end{eqnarray}
Further, after doing the $T$ integral by steepest descents, one reproduces exactly  the WKB result  \eqn{wkb-1}, including the prefactor term \cite{wli2}.

This agreement with previous one dimensional semiclassical results is encouraging, but the real advantage of the worldline instanton method is that it can be extended to more general fields. 
Technically, there are  several steps to the computation: (i) find 
 periodic solutions [the worldline instantons] to the Euclidean classical equations of motion \eqn{euler}
; (ii)
The dominant exponential factor in the pair production rate is $e^{-S[x^{\rm cl}]}$, involving the classical action evaluated on the worldline instanton path; (iii)
The prefactor coming from the semiclassical approximation to the quantum mechanical path integral is given by the determinant of the fluctuation operator in (\ref{morse}), evaluated on the worldline instanton path; (iv) 
The $T$ integration fixes the conserved quantity $\frac{1}{4}\,\dot{x}_\mu^2$ to take the value $m^2$, and also produces a prefactor; (v)
Possible residual dependence on the spacetime location of the loop must be integrated over. In a Gaussian approximation this also leads to another prefactor.

\subsection{Worldline Instantons and the Gutzwiller Trace Formula}

Note that in general there are three apparently separate 
[but below we shall see that they are in fact related!] 
prefactor contributions to the final answer: one coming from each integration in \eqn{feynman}.
This procedure is reminiscent of that used to derive the Gutzwiller trace formula \cite{gutzwiller,littlejohn,cvitanovic} for the trace of the Green's function in non-relativistic quantum mechanics:
\begin{eqnarray}
{\rm tr}\, G(E)&=&\int d^3x\,\, \int_0^\infty dt\,\, e^{\frac{i}{\hbar} E\, t}\,\,\langle x\,|\, e^{\frac{i}{\hbar}\, H\,t}\,|\, x\rangle 
\label{gutzwiller1}\\
&=&\sum_{\rm orbits\,\, p}\, T_p\, \frac{e^{\frac{i}{\hbar} S_p(E)-i\frac{\pi}{2}\,m_p}}{\sqrt{\det(1-{\bf J}_p)}}
\quad .
\label{gutzwiller2}
\end{eqnarray}
Here $T_p$ is the period of the ${\rm p}^{\rm th}$ classical closed orbit of energy $E$, $S_p(E)$ is its action, $J_p$ is the associated monodromy matrix [see below], and $m_p$ are Maslov indices.  Before proceeding to the corresponding worldline instanton expression, we briefly sketch the strategy of the derivation of the Gutzwiller trace formula [an excellent introduction for physicists is in \cite{cvitanovic}]. The first step is to make a semiclassical  approximation for the propagation kernel in  \eqn{gutzwiller1} between distinct points $x$ and $x^\prime$, in a time interval $t$:
\begin{eqnarray}
\langle x\,|\, e^{\frac{i}{\hbar}\, H\,t}\,|\, x^\prime\rangle \approx \sqrt{\det\,\left |\frac{\partial^2 R}{\partial x\,\partial x^\prime}\right |}\,\, e^{\frac{i}{\hbar}\, R(x,\, x^\prime;\, t)}
\label{gsemi}
\end{eqnarray}
Here, $R(x, x^\prime; t)$ is Hamilton's principal function for the path connecting the points $x$ and $x^\prime$, in time $t$. The prefactor is the Van Vleck determinant factor, given by variations with respect to the path's endpoints. 
The next step is to evaluate the $t$ integral in \eqn{gutzwiller1} by the stationary phase approximation. Here we note that the exponent is $\frac{i}{\hbar}[E\, t+R(x,\, x^\prime;\, t)]$, whose stationary point condition, $\frac{\partial R}{\partial t}=-E$, fixes $t$ to be the period such that the closed classical path has energy equal to $E$. This means that the stationary point condition implements the Legendre transform from $R(x,\, x^\prime;\, t)$ to the action $S(x, x^\prime; E)$ associated with the closed classical orbit of energy $E$. Furthermore, the prefactor from the $t$ integral contributes a factor $1/\sqrt{-\frac{\partial^2 R}{\partial t^2}}=\sqrt{\frac{\partial^2 S}{\partial E^2}}$, 
which follows from the other Legendre transform relation $\frac{\partial S}{\partial E}=t$.
The final step towards the Gutzwiller trace formula (\ref{gutzwiller2}) is to take the coincident point limit, $x^\prime\to x$, and to integrate over the marked point $x$, the beginning and end of the orbit. Evaluating this integral by a third and final steepest descents approximation forces the closed loop to be periodic. The $x$ integral naturally splits into an integral along the loop, and an integral transverse to the loop. The integral along the orbit produces a factor of the period $T_p$, reflecting translation invariance with respect to the parameter of the orbit. In the language of quantum field theory, this is a collective coordinate contribution, associated with the zero mode coming from the basic fact from classical mechanics that the fluctuation operator has a zero mode: $\Lambda_{\mu\nu}\dot{x}^{\rm cl}_\nu=0$. On the other hand, the transverse $x$ integrations give yet another determinant prefactor $1/\sqrt{\det \left | \frac{\partial^2 S}{\partial x_\perp \partial x_\perp^\prime}\right |}$. 

The remarkable result of Gutzwiller \cite{gutzwiller} is that these three different-looking prefactors [coming successively from semiclassical approximations to the quantum mechanical path integral, the $t$ integral, and finally the $x$ trace] all combine into a {\it single determinant prefactor} that has a simple geometrical interpretation in phase space:
\begin{eqnarray}
\hskip -.4cm \int dx \left[
\frac{\sqrt{\frac{\partial^2 R}{\partial x\,\partial x^\prime}}\,\sqrt{\frac{\partial^2 S}{\partial E^2}}}
{\sqrt{\frac{\partial^2 S}{\partial x_\perp \partial x_\perp^\prime}}}\,e^{\frac{i}{\hbar} S(x,\,x^\prime;\, E)}\right]_{x=x^\prime}  = \frac{T_p\, e^{\frac{i}{\hbar}\, S[E]}}{\sqrt{{\rm det}\left({\bf 1}-{\bf J}\right)}}
\label{gut}
\end{eqnarray}
Here ${\bf J}$ is the monodromy matrix of the periodic orbit having energy $E$ and period $T_p$. For a precise definition of ${\bf J}$ see \cite{gutzwiller,littlejohn,cvitanovic}, including the Maslov indices, which we have suppressed here for the sake of simplicity. It is sufficient here to note that the full prefactor is expressed in terms of a classical invariant of the closed orbit, which characterizes the behavior of small deviations from the periodic orbit in phase space.
\begin{figure}[h]
\centerline{\resizebox{0.4\textwidth}{!}{%
\includegraphics{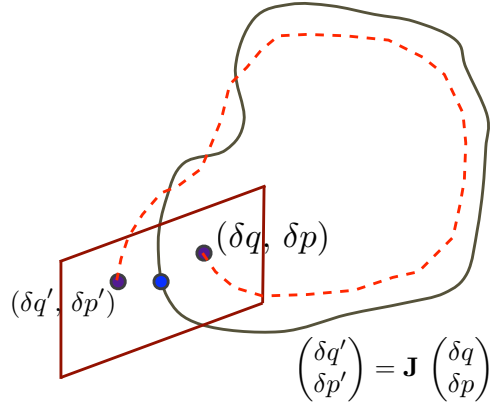}}
}
\caption{A pictorial representation of the geometrical meaning of the monodromy matrix ${\bf J}$ appearing in expressions (\ref{gut}) and (\ref{final}). We consider  closed orbit [solid, grey line] in phase space, and displace the starting point slightly in the plane transverse to the loop, and then propagate according to the equations of motion, returning after one period to another point $(\delta q^\prime, \delta p^\prime)$ in the same transverse plane. The monodromy matrix relates the starting and end points: $(\delta q^\prime, \delta p^\prime)={\bf J}\,(\delta q, \delta p)$.}
\label{fig9}
\end{figure}

These steps we have just sketched for the derivation of Gutzwiller's trace formula are closely analogous to the steps required in evaluating the worldline path integral expresion \eqn{feynman} for the effective action $\Gamma[A]$. Thus, analogous to \eqn{gutzwiller2}, we seek the following representation of the imaginary part of the effective action:
\begin{eqnarray}
{\rm Im}\, \Gamma&=&{\rm Im}\,\int d^4x\,\, \int_0^\infty \frac{dT}{T}\,\, e^{-m^2\, T}\,\,\langle x\,|\, e^{-(-D^2_\mu)\, T}\,|\, x\rangle
\nonumber\\
&=&\sum_{\rm orbits\,\, p}\, \frac{e^{- S_p(m^2)}}{\sqrt{\det(1-{\bf J}_p)}}
\end{eqnarray}
The effective action $\Gamma[A]$ in (\ref{feynman}) involves a Euclidean path integral, in four dimensional space rather than a path integral in three dimensional nonrelativistic quantum mechanics. There  is also an extra factor of $1/T$ because $\Gamma[A]$ is a log determinant. Nevertheless, the ideas follow through, and in the end we can combine the three prefactors found by following the worldline instanton strategy enumerated above, to form one simple prefactor expressed in terms of the monodromy matrix for the periodic orbit, now viewed in phase space  \cite{dd}. The successive approximations are first for the
the quantum mechanical path integral:
\begin{eqnarray}
K(x, x^\prime; T)&:=&\langle x| e^{-T(-D_\mu^2)}\, |x^\prime\rangle 
\label{semi}\\
&\approx& \frac{1}{(2 \pi)^2}\,
\sqrt{\left| \det\left(\frac{\partial^2 R}{\partial x\, \partial x^\prime}\right)\right |}\, e^{-R(x, x^\prime; T)}
\nonumber
\end{eqnarray}
where $R(x, x^\prime; T)$ is the Hamilton principal function for the classical trajectory  from $x$ to $x^\prime$ in four-dimensional Euclidean
space, in the proper-time interval $T$. 
Next, the $T$ integral is evaluated by steepest descents. The critical point of the exponential factor arises when $\frac{\partial R}{\partial T}=-m^2$. This has a natural classical interpretation in terms of the Legendre transformation between the Hamilton principal function $R(x, x^\prime; T)$ [expressed in terms of the total time elapsed along the trajectory] and the action $S(x, x^\prime; E)$ [expressed in terms of the constant energy of the trajectory]: $R(x, x^\prime; T)=S(x, x^\prime; E)-E\, T$.
It follows that $\frac{\partial R}{\partial T}=- E$, and $ \frac{\partial S}{\partial E}=T$.
Thus,  the critical point $T_c$ of the $T$ integral occurs when $E=m^2$, so that 
\begin{eqnarray}
 \int_0^\infty \frac{dT}{T}e^{-m^2 T} K(x, x^\prime; T)\approx  \frac{1}{(2 \pi)^2\, T_c}\, \sqrt{\left| \det\left(\frac{\partial^2 R}{\partial x\, \partial x^\prime}\right)\right
|_{T_c}}\, \sqrt{\frac{2\pi}{\left | \frac{\partial^2 R}{\partial T^2}\right |_{T_c}}}\, e^{-S(x, x^\prime; m^2)}
 \label{tint}
\end{eqnarray}
The final step is the coincident limit $x\to x^\prime =x^{(0)}$, and the trace over $x^{(0)}$, yielding
\begin{eqnarray}
{\rm Im} \, \Gamma\approx \frac{e^{-S(E=m^2)}}{\sqrt{\det\left({\bf
1}-{\bf J}\right)}}\quad . 
\label{final}
\end{eqnarray}
The main advantage of expressing the computation in this language of the Gutzwiller trace formula  is that the total prefactor is encapsulated in a {\it single determinant}, which moreover has a natural mathematical and geometrical meaning
in the Euclidean phase space \cite{dd}. In previous work \cite{brezin,popov-wkb,kimpage,wli2,wli3}  the various prefactor contributions
have been evaluated separately, and then combined at the end. Thus, the  computational strategy of the worldline instanton approach can be stated succinctly as follows:
\begin{enumerate}
\item solve the classical equations of motion in four dimensional Euclidean space to
find all closed periodic trajectories of "energy" $E\equiv\frac{\dot{x}^2}{4}=m^2$: such a solution is a ``worldline instanton''.
\item evaluate the classical action $S_{\rm cl}(E=m^2)$ on these trajectories. Then the leading exponential contribution to ${\rm Im}\,\Gamma$ is $e^{-S_{\rm cl}}$.
\item compute the prefactor from the monodromy matrix ${\bf J}$ for the dominant trajectory(ies).
\end{enumerate}

As an illustration of this procedure, consider again the case of a one-dimensional inhomogeneity,  for example, the case of a time dependent electric field directed in the $x_3$ direction. As noted above, for such a configuration the vacuum pair production probability can be computed in several other ways using WKB methods \cite{brezin,popov-wkb,kimpage,wli2}, and so it is possible to make a direct comparison.
We can choose a Euclidean gauge field
$
A_3(x_4)=\frac{\mathcal E}{\omega}\, f(\omega\, x_4)
$,
where ${\mathcal E}$ characterizes the overall magnitude of the associated electric field, $\omega$ characterizes the scale of the time dependence, and $f(\omega\, x_4)$ is some smooth function. For example, for a constant electric field ${\mathcal E}(t)={\mathcal E}$, we have $f(x)=x$; for a sinusoidal electric field ${\mathcal E}(t)={\mathcal E}\,
\cos(\omega\, t)$, we have $f(x)=\sinh(x)$; and for a single-pulse electric field ${\mathcal E}(t)={\mathcal E}\, {\rm
sech}^2(\omega\, t)$, we have $f(x)=\tan(x)$. Then the classical action on a periodic trajectory of energy $E$ can be written [here $y:=\frac{e {\mathcal E}}{\omega \sqrt{E}}\, f(x)$]
\begin{eqnarray}
S(E)&=&\oint d x_4 \sqrt{E-\left(\frac{e {\mathcal E}}{\omega}\, f(\omega \, x_4)\right)^2}\nonumber\\
&=&\frac{2E}{e\, {\mathcal E}}\int_{-1}^1 dy\,
\frac{\sqrt{1-y^2}}{\left[f^\prime(z) \right]_{z=z(y)}}
\end{eqnarray}
Evaluated at $E=m^2$, this is precisely the exponent, $g(\gamma^2)\,m^2\pi/(e{\mathcal E})$, appearing in the standard WKB result for the pair
production rate \cite{brezin,popov-wkb,kimpage,wli2}.
To evaluate the prefactor, we can choose $x_4$ as $x_\parallel$. Then the transverse $x_3$ direction is in fact an invariant "flat" direction, so we do not need to perform the transverse integration. 
This illustrates  the important point that \eqn{final} must be interpreted appropriately when there are physical zero modes associated with symmetries of the field.
Thus, we go back to 
\eqn{tint} and observe that $\frac{\partial^2 R}{\partial T^2}=-1/\frac{\partial^2 S}{\partial E^2}$. Furthermore, the other determinant factor in \eqn{tint}
is easily computed (see \cite{wli2}) using the Gel'fand-Yaglom formula:
\begin{eqnarray}
\left . \det\left(\frac{\partial^2 R}{\partial x\, \partial x^\prime}\right)\right |_{x=x^\prime}=
\frac{m^4}{16\, E^3 \, T^2}\,
\frac{1}{\dot{x}_4^2 \left(\frac{\partial^2 S}{\partial E^2}\right)^2}\quad .
\end{eqnarray}
Thus, relative to the constant spatial volume $V_3$,
\begin{eqnarray}
\frac{{\rm Im} \, \Gamma^{\rm WLI}}{V_3}\approx \frac{\sqrt{2\pi}}{2(4\pi)^2 m} \left[ \frac{e^{-S(E)}}{\frac{\partial S}{\partial E}\sqrt{\frac{\partial^2 S}{\partial E^2}}} \right]_{E=m^2}\quad ,
\label{1dfinal}
\end{eqnarray}
Note that (\ref{1dfinal}) agrees precisely with the conventional WKB result in \eqn{wkb-1}, with the added semiclassical interpretation of the various terms.

\section{Conclusions}

I conclude by reiterating that the ELI project has great promise to open up an entirely new non-perturbative regime of QED, and of quantum field theories in general. There are many experimental and theoretical challenges ahead, but the future appears bright. Theoretically, the biggest challenge in the non-perturbative arena is to develop efficient techniques, both analytical and numerical, for computing the effective action and related quantities, in  external fields that realistically represent the experimental laser configurations. A lot of progress has been made in this direction, but new ideas and methods are still needed. Another important issue is that of back-reaction, which is typically ignored in the limit where the electric field strength is below the critical field in (\ref{critical}). But the combination of an ELI laser with an accelerated electron beam may produce effective field strengths well above the critical field, in which case back-reaction becomes significant. Insights from these nonperturbative QED studies should have implications for related questions in gravitational systems of quantum field theory in strongly curved spaces, and also for non-pertubative effects in nonabelian gauge theories \cite{dima}.

\vskip .5cm

{\bf Acknowledgements:} I thank the organizers of the ELI Frauenw\"orth conference, especially D. Habs and M. Gross, for an extremely interesting meeting, and I thank G. Mourou and T. Tajima for helpful discussions. I acknowledge support from the DOE through the grant DE-FG02-92ER40716,  and from the European Commission under contract ELI pp 212105 in the framework of the program FP7 Infrastructures-2007-1.

%
%

\end{document}